\documentclass[11pt,onecolumn,draftcls]{IEEEtran}

\usepackage{graphicx,amsmath}
\usepackage[caption=false,font=footnotesize]{subfig}

\begin{document}

\title{Interactive Reconciliation with Low-Density Parity-Check Codes}

\author{
  \IEEEauthorblockN{
    Jes\'us~Mart\'inez-Mateo,
    David~Elkouss
    and~Vicente~Mart\'in
  }\\
  \IEEEauthorblockA{
    Research group on Quantum Information and Computation\\
    Universidad Polit\'ecnica de Madrid (UPM)\\
    Campus de Montegancedo, 28660 Boadilla del Monte (Madrid), Spain\\
    e-mail: \{jmartinez, delkouss, vicente\}@fi.upm.es
  }
}

\maketitle

\begin{abstract}

Efficient information reconciliation is crucial in several scenarios, being quantum key distribution a remarkable example. However, efficiency is not the only requirement for determining the quality of the information reconciliation process. In some of these scenarios we find other relevant parameters such as the interactivity or the adaptability to different channel statistics. We propose an interactive protocol for information reconciliation based on low-density parity-check codes. The coding rate is adapted in real time by using simultaneously puncturing and shortening strategies, allowing it to cover a predefined error rate range with just a single code. The efficiency of the information reconciliation process using the proposed protocol is considerably better than the efficiency of its non-interactive version.

\end{abstract}


\section{Introduction}

Since the publication of the first quantum protocol, more than 25 years ago, quantum key distribution (QKD)~\cite{Gisin02} has evolved into a functional and commercial technology, and nowadays it is already possible to find commercial QKD systems by several manufacturers. A QKD system is used to create secret keys between two parties connected through a quantum channel, i.e. for instance an optic fibre. However, this technology is still far from reaching its real potential due to the lack of suitable developments in some of its fundamental processes, such as error correction. In a QKD protocol, error correction is included within a broader process known as secret key distillation~\cite{Maurer93}. In this process, error correction is a procedure used to reconcile discrepancies between two bit sequences, for this reason this procedure is known as information reconciliation. In order to accomplish it, the parties must exchange additional information over a public but authenticated channel: it can be read but not modified by an hypothetical eavesdropper. Since the information exchanged for reconciliation provides information about the key, the parties must agree on an additional procedure, called privacy amplification~\cite{Bennett95}, used to reduce the information that may have been derived by any eavesdropper. An optimal reconciliation procedure provides the minimum information required for correcting the discrepancies between two sequences, minimising the key material that must be discarded during the privacy amplification, therefore maximising the final secret-key length.

One of the first methods proposed for correcting errors in a QKD system was \textit{Cascade}~\cite{Brassard94}. Currently, it is probably the most widely used procedure for this purpose in QKD, due to its simplicity and relatively good efficiency (see Fig.~\ref{fig:efficiency}). However, \textit{Cascade} is a highly interactive process that requires many communication rounds. The parties have to exchange a large number of messages where parities of different blocks and subblocks of a key are published.


A better alternative for error correction in QKD systems is provided by other strategies such as low-density parity-check (LDPC) codes. These codes were introduced by Gallager in the early 60s~\cite{Gallager62}, and recently several proposals have emerged for using LDPC codes in the information reconciliation process~\cite{Elkouss09, Elkouss10}. In this paper we propose a new protocol for error correction using rate adaptive LDPC codes. The protocol is able to correct errors within a known error rate range, iteratively transmitting more symbols in order to minimise the information transmitted for correction.

\begin{figure}
\centering
\includegraphics[width=0.6\linewidth]{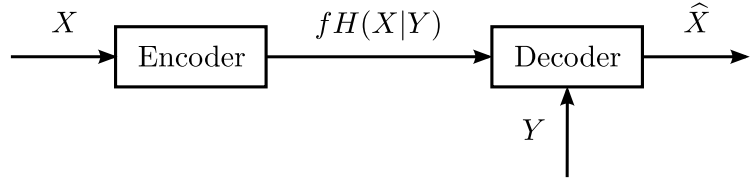}
\caption{Source coding with side information for one-way reconciliation.}
\label{fig:side-information}
\end{figure}

The paper is organised as follows: First, in section~\ref{sec:reconciliation}, it is described the problem of information reconciliation in the secret-key agreement context. Then, in section~\ref{sec:protocol}, a new protocol is proposed to improve the reconciliation process using interactive communication between the parties. Finally, in section~\ref{sec:results}, results with this new protocol are shown. These results are also compared with two different approaches: a similar proposal using rate adaptive codes but without interactive communication, and the simplest approach using LDPC codes without rate modulation.

\begin{figure*}[!t]
\centering
\subfloat[Puncturing]{\includegraphics[width=0.5\linewidth]{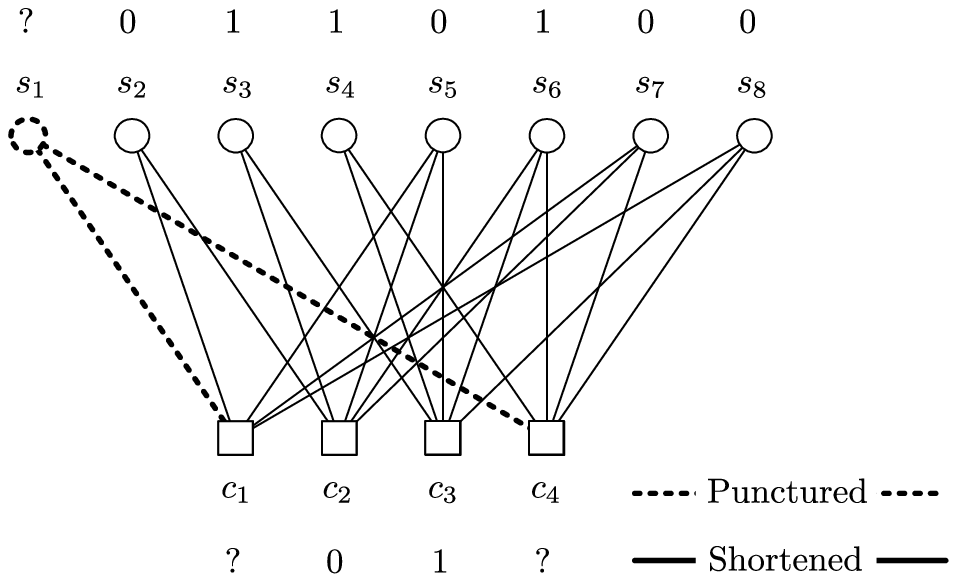}
\label{fig:puncturing}}
\hfil
\subfloat[Shortening]{\includegraphics[width=0.41\linewidth]{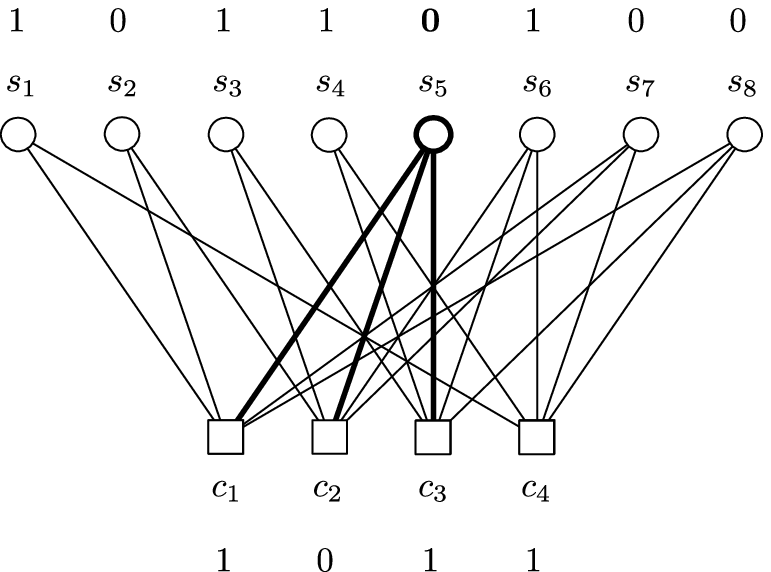}
\label{fig:shortening}}
\caption{Examples of puncturing and shortening strategies applied to an LDPC code in order to modulate the coding rate.}
\label{fig:ps}
\end{figure*}

\section{Information Reconciliation}
\label{sec:reconciliation}

The problem of information reconciliation, when only one-way transmissions are allowed, can be modelled by the more general problem of source coding with side information. In this section we describe this more general approach, and are reviewed those techniques used to adapt LDPC codes in the information reconciliation context.

\subsection{Source Coding with Side Information}

Let $X$ and $Y$ be two discrete random variables representing two correlated sources, and let $X^n$ and $Y^n$ be two correlated sequences obtained from both sources respectively. Assuming that these sources are separated into two legitimate parties: Alice and Bob. Information reconciliation allows Bob to recover $X^n$ with the help of $Y^n$ and sending $M$ messages over a lossless channel. 
In the source coding with side information description, one of the parties encode the sequence $X^n$, and the other recovers $\hat{X}^n$ using the information provided by the encoded sequence and with help of side information $Y^n$, such that $X^n = \hat{X}^n$ with high probability. The minimum rate for encoding the source $X$ in order to get $X=\hat{X}$ with the side information provided by $Y$ was determined by Slepian-Wolf to be $H(X|Y)$~\cite{Slepian73} (see Fig.~\ref{fig:side-information}). Both problems, information reconciliation and source coding with side information, are equivalent if only one-way transmissions are allowed, and $H(X|Y)$ is the minimum rate that could be used in a reconciliation protocol. However, even though the problem is formally different, an interactive reconciliation process shares the same lower bound~\cite{VanAssche06}. Thus the reconciliation efficiency, $f$, for both one-way and interactive protocols can be defined by:

\begin{equation}
\label{eq:efficiency}
f = \frac{R_X}{H(X|Y)} \geq 1
\end{equation}

In the quantum cryptography context information reconciliation arise after the basis reconciliation process, this is when both parties of a QKD system, Alice and Bob, share a raw key with discrepancies that should be removed by following a key distillation process. In most of QKD protocols, e.g. BB84~\cite{Bennett84} or SARG~\cite{Scarani04}, these discrepancies are uncorrelated and symmetric, such that they can be interpreted as errors in a communication made through a binary symmetric channel (BSC).

\subsection{LDPC Codes and Syndrome Decoding}

LDPC codes are known to achieve coding rates near the capacity of several channels under belief propagation decoding~\cite{Richardson01}. It has been also shown that these codes can be used to encode near the theoretical limit for source coding with side information~\cite{Liveris02}. A modified decoder was proposed for syndrome decoding and applying the bin approach by Wyner~\cite{Zamir02}. The use of LDPC codes for encoding correlated sources was later formalised~\cite{Muramatsu05}.

Following this exposition, information reconciliation can be solved for many QKD protocols by using good LDPC codes for the BSC. This problem has been already addressed, and good families of these codes have been found for different coding rates~\cite{Elkouss09}. However, an LDPC code is constructed for a fixed coding rate. In consequence, in those scenarios of varying characteristics, such as QKD, if the parties do not share a suitable number of codes, the efficiency curve shows a saw behaviour (see Fig.~\ref{fig:results}). In order to solve this behaviour, in the next section we describe a new protocol able to adapt the coding rate of an LDPC code, minimising the information revealed for reconciliation.

\section{Protocol}
\label{sec:protocol}

\subsection{Rateless Coding}

Puncturing and shortening are two suitable strategies able to adapt the rate of a channel code as we have already shown in a previous work~\cite{Elkouss10}. When $p$ punctured symbols of a codeword are removed, a $[n, k]$ code is converted into a $[n-p, k]$ code (see Fig.~\ref{fig:puncturing}). Whereas, when shortening, $s$ symbols are removed during the encoding process, and a $[n, k]$ code is converted into a $[n-s, k-s]$ code (see Fig.~\ref{fig:shortening}).

Supposing that $R_0$ is the original coding rate of a family of LDPC codes defined by:

\begin{equation}
R_0 = 1 - \frac{\displaystyle \sum_i \lambda_i / i}{\displaystyle \sum_j \rho_j / j}
\end{equation}

\noindent where $\lambda_i$ and $\rho_i$ are the coefficients of their generating polynomials. This rate can be modulated applying puncturing and shortening procedures as defined below. The modulated rate is then calculated as:

\begin{equation}
R = \frac{k - s}{n - p - s} = \frac{R_0 - \sigma}{1 - \pi - \sigma}
\end{equation}

\noindent where $\pi = p/n$ and $\sigma = s/n$ are the ratios of punctured symbols and shortened symbols respectively.

Both strategies, puncturing and shortening, may be applied in an isolated way in order to increase or decrease the coding rate respectively. However, we propose the use of both strategies simultaneously defining a new constant parameter, $\delta = \pi + \sigma$. This proposal is based primarily on two reasons:

\begin{enumerate}

\item Applying the same proportion of puncturing and shortening in every modulation. In consequence, regardless of the coding rate, the key length that can be corrected using the modulated code is known in advance.

\item As it is discussed in the next section, the use of puncturing and shortening simultaneously allow us to modify a previously modulated code in order to decrease the coding rate and to repeat an unsatisfactory correction process. 

\end{enumerate}
 
\begin{figure}
\centering
\includegraphics[width=0.7\linewidth]{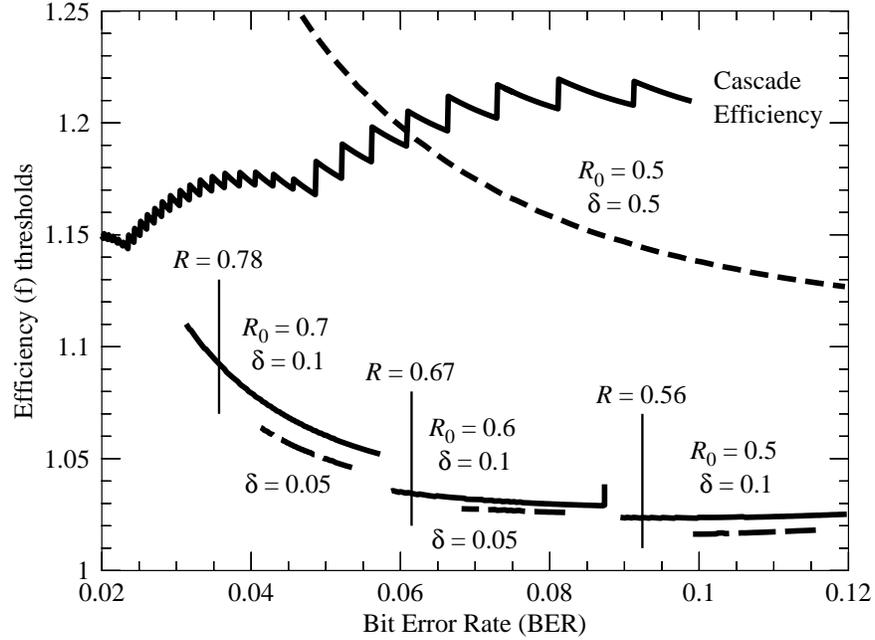}
\caption{Efficiency of \textit{Cascade}~\cite{Brassard94} and efficiency thresholds of LDPC codes with rate modulation for an error rate from 2\% to 10\%. The range is representative of good to very bad quantum channels for the QKD case. Efficiencies have been calculated, using the expression defined in Eq.~\ref{eq:efficiency}, for three different codes with rates $R_0$: $0.5$, $0.6$ and $0.7$. Two $\delta$ values, $0.1$ and $0.05$, have been used for the rate modulation of these codes. An additional $\delta = 0.5$ has been used with $R_0 = 0.5$ covering the entire error rate range. The curves show how the efficiency of the LDPC code depends on the ratio of puncturing and shortening, $\delta$, and the original coding rate, $R_0$. Higher $\delta$ values imply a bigger range of coding rates covered, unfortunately the efficiency drops for high values of $\delta$~\cite{Pishro-Nik07}. The efficiency drop increases for simultaneous use of high $\delta$ values with high coding rates.}
\label{fig:efficiency}
\end{figure}

The efficiency, defined in Eq.~\ref{eq:efficiency}, of the modulated code depends on the coding rate and the ratio $\delta$ of puncturing and shortening as shown in Fig.~\ref{fig:efficiency}.

\subsection{Interactive Reconciliation}

\begin{figure}
\centering
\includegraphics[width=0.6\linewidth]{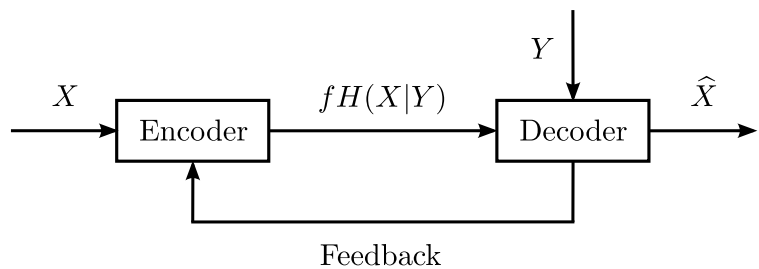}
\caption{Source coding with side information and feedback.}
\label{fig:interactive-decoding}
\end{figure}

The process of adapting the coding rate of an LDPC code is usually done with a previous estimate of the error rate to be corrected, this estimation is traditionally carried out by exchanging a sample of the sequence on the public channel. We propose here a new protocol for information reconciliation with LDPC codes that does not require to estimate the error rate. The protocol is based on syndrome coding, but adding a new functionality: feedback information about the success of the decoding process (see Fig.~\ref{fig:interactive-decoding}). With this feedback the original one-way approach becomes an interactive protocol, as described below, with more flexibility in order to correct optimally a range of error rates. The protocol is blind in the sense that it is able to adapt to different channel configurations without a prior estimate and, the reconciliation is successful as long as the channel's characteristics are within a pre-established range.

The protocol is described by the following three steps:

\subsubsection*{Step 0) Raw Key Exchange}

Initially it is assumed that two sources, $X$ and $Y$, generate two correlated symbol sequences, $\mathbf{x}$ and $\mathbf{y}$ belonging to Alice and Bob respectively. Moreover, it is also supposed that the two symbol sequences have discrepancies within a bounded error rate range, $[e_0, e_1]$. From this hypothesis, Alice and Bob can choose an LDPC code with an information rate aimed to correct an intermediate point in the interval. Depending on the range and the efficiency target, the parties agree on a $\delta$ value to cover the entire range of required coding rates.

\begin{equation}
R_\mathrm{min} = \frac{R_0 - \delta}{1 - \delta} \leq R \leq \frac{R_0}{1 - \delta} = R_\mathrm{max}
\end{equation}

\noindent such that $R_\mathrm{min} \leq 1 - h_2(e_1)$ and $R_\mathrm{max} \geq 1 - h_2(e_0)$, where $h_2$ is the binary Shannon entropy.

As initial coding rate is chosen the highest value, $R = R_\mathrm{max}$, such that all symbols used to modulate the rate correspond to a punctured symbol, i.e. $\delta = \pi$ and $\sigma = 0$. In this case the protocol provides the minimum amount of information.

\subsubsection*{Step 1) Encoding}

Once it has been established a value for the coding rate, both parties compute the number of symbols to be punctured and shortened, $p$ and $s$ respectively:

\begin{equation}
\begin{split}
s & = \lceil \left( R_0 - R (1 - \delta) \right) n \rceil \\
p & = \lfloor \delta n \rfloor - s
\end{split}
\end{equation}

The first time this step is run, Alice randomly chooses the symbols to be punctured ---there are no shortened symbols in the first round---, and set them with random values. Once Alice knows which positions correspond with punctured symbols, and their values, she calculates the syndrome (compressed information), $\mathbf{z} = \mathbf{x} H^t$, and sends it to Bob along with their positions.

In subsequent runs of this step, Alice chooses randomly a preestablished proportion of punctured symbols that will be converted to shortened symbols and transmits to Bob their positions and values. The proportion of converted symbols in each round must be agreed by both parties at the beginning of the protocol, and it depends on the maximum number of rounds that is allowed and the desired efficiency.

\subsubsection*{Step 2) Decoding}

Bob uses his correlated sequence, $\mathbf{y}$, and the information provided by punctured and shortened symbols as starting point to find a sequence with a syndrome that matches the syndrome received from Alice, $\mathbf{z}$. The protocol is successfully concluded in this step when Bob decodes a word matching the syndrome received. Otherwise, if the decoding process is stopped because the maximum number of pre-set iterations has been reached, then Bob agrees with Alice a decrease in the coding rate, if possible, and they return to the previous Step~1.

The protocol fails if the coding rate takes its minimum value and decoding is unsuccessful, i.e. $R = (R_0 - \delta) / (1 - \delta)$, which happens for $\delta = \sigma$ and $\pi = 0$.

\begin{figure}
\centering
\includegraphics[width=0.6\linewidth]{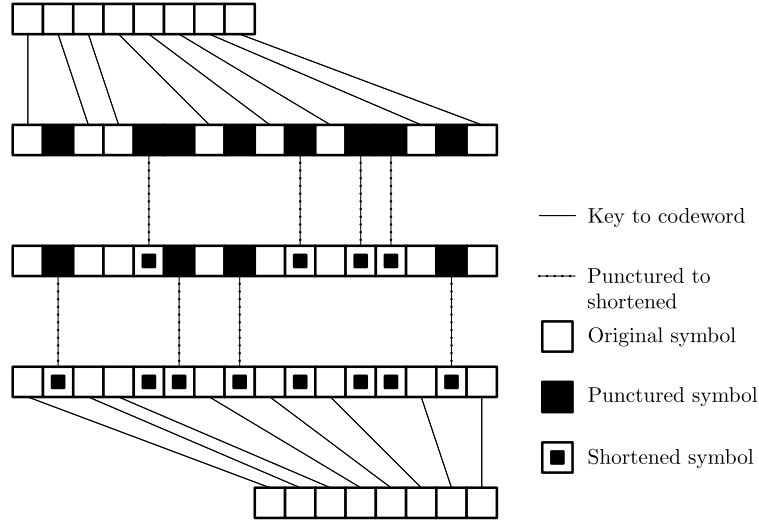}
\caption{Example of a complete execution of the interactive protocol. The example shows how the symbols of an initial sequence are distributed in different positions of a codeword. The remaining positions of the codeword are initially marked as punctured symbols. In each round, the protocol replaces a proportion of punctured symbols with shortened symbols, thus reducing the coding rate.}
\label{fig:protocol}
\end{figure}

A graphic description of this protocol is shown in Fig.~\ref{fig:protocol}. The figure illustrates an example showing three executing rounds.

Different executions of the proposed protocol may conclude with different ratios for puncturing and shortening, $\pi$ and $\sigma$ respectively, i.e. different protocol executions reconcile the original sequences with different efficiencies. Efficiency for a single protocol execution is defined in Eq.~\ref{eq:efficiency}, the efficiency of this protocol can be measured by taking an average value. Puncturing and shortening ratios are then calculated by:

\begin{equation}
\hat{\pi} = \frac{1}{M} \sum_{i = 1}^{M} \pi^{(i)}
\; ; \quad
\hat{\sigma} = \frac{1}{M} \sum_{i = 1}^{M} \sigma^{(i)}
\end{equation}

\noindent where $M$ is the number of executions. The average efficiency is then calculated as:

\begin{equation}
\label{eq:average-efficiency}
\hat{f} = \frac{1}{h_2(e)} - \frac{R_0 - \hat{\sigma}}{(1 - \delta) h_2(e)} = \frac{1 - R_0 - \hat{\pi}}{(1 - \delta) h_2(e)}
\end{equation}

\noindent where $e$ is the crossover probability that has been corrected.

Let $Q$ be the maximum number of rounds, the efficiency value in an isolated execution is increased in each round by a constant factor, $\epsilon$, that depends on the proportion of new shortened symbols, $q = \delta / Q$ such that $\pi_{j + 1} = \pi_{j} - q$ and $\sigma_{j + 1} = \sigma_{j} + q$. The efficiency of an execution that concludes in the round $j$ can be also calculated by $f_{j} = f_{0} + j \epsilon$, where:

\begin{equation}
f_{0} = \frac{1 - R_0 - \delta}{(1 - \delta) h_2(e)}
\; ; \quad
\epsilon = \frac{q}{(1 - \delta) h_2(e)}
\end{equation}

\section{Results}
\label{sec:results}

In order to produce a representative set of simulations that demonstrate the operation of the proposed protocol, we decided to build a single LDPC code of length $n = 2 \times 10^5$ and rate $R_0 = 0.6$ using a family of codes proposed by Elkouss~\cite{Elkouss09} for the BSC. The code, with a puncturing and shortening proportion of $\delta = 0.1$, is able to modulate coding rates from $R_\mathrm{min} = 0.56$ to $R_\mathrm{max} = 0.67$, i.e. it is possible to construct codes for correcting error rates in an approximate range from 6\% to 9\%. Table~\ref{tab:rounds} shows the proportion and number of punctured and shortened symbols in each round, assuming that only a maximum of seven rounds can be executed. For convenience, the proportions of punctured and shortened symbols has been normalised to 1, such that $\pi = \delta \pi^*$ and $\sigma = \delta \sigma^*$.

\begin{table}
\caption{Proportion and number of punctured and shortened symbols, and modulated rate per round.}
\label{tab:rounds}
\centering
\begin{tabular}{|c|c|c|c|r|r|c|c|c|}
\hline
Round & $\delta$ & $\pi^*$ & $\sigma^*$ & \multicolumn{1}{|c|}{$p$} & \multicolumn{1}{|c|}{$s$} & $R_0$ & $R$ \\
\hline\hline
0 & 0.1 & 1.00 & 0.00 & 20000 &     0 & 0.6 & 0.67 \\
1 & 0.1 & 0.83 & 0.17 & 16666 &  3334 & 0.6 & 0.65 \\
2 & 0.1 & 0.67 & 0.33 & 13332 &  6668 & 0.6 & 0.63 \\
3 & 0.1 & 0.50 & 0.50 &  9999 & 10001 & 0.6 & 0.61 \\
4 & 0.1 & 0.33 & 0.67 &  6666 & 13334 & 0.6 & 0.59 \\
5 & 0.1 & 0.17 & 0.83 &  3333 & 16667 & 0.6 & 0.57 \\
6 & 0.1 & 0.00 & 1.00 &     0 & 20000 & 0.6 & 0.56 \\
\hline
\end{tabular}
\end{table}

Table~\ref{tab:results} shows the average number of rounds, $\hat{N}$, needed for correcting different error rates using a previously constructed LDPC code. The table also includes the average number of punctured symbols, $\hat{p}$, and the average number of shortened symbols, $\hat{s}$, that have been used in the last round of the correction process. From the average number of punctured and shortened symbols, the average proportion of punctured and shortened symbols, $\hat{\pi}$ and $\hat{\sigma}$ respectively, have been calculated together with the average efficiency, $\hat{f}$, as defined in Eq.~\ref{eq:average-efficiency}.

\begin{table}
\caption{Average number of rounds needed to correct different error rates with a single but modulated LDPC code.}
\label{tab:results}
\centering
\begin{tabular}{|c|c|r|r|c|c|c|}
\hline
BER & $\hat{N}$ & \multicolumn{1}{|c|}{$\hat{p}$} & \multicolumn{1}{|c|}{$\hat{s}$} & $\hat{\pi}$ & $\hat{\sigma}$ & $\hat{f}$ \\
\hline\hline
0.055 & 0.03 & 19900 &   100 & 0.0995 & 0.0005 & 1.08664 \\
0.060 & 1.12 & 16266 &  3734 & 0.0813 & 0.0187 & 1.08144 \\
0.065 & 2.36 & 12132 &  7868 & 0.0607 & 0.0393 & 1.08651 \\
0.070 & 3.12 &  9599 & 10400 & 0.0480 & 0.0520 & 1.06883 \\
0.075 & 4.38 &  5399 & 14601 & 0.0270 & 0.0730 & 1.07841 \\
0.080 & 5.00 &  3333 & 16667 & 0.0167 & 0.0833 & 1.05895 \\
0.085 & 6.00 &     0 & 20000 & --     & --     & -- \\
0.090 & 6.00 &     0 & 20000 & --     & --     & -- \\
\hline
\end{tabular}
\end{table}

\begin{figure}
\centering
\includegraphics[width=0.6\linewidth]{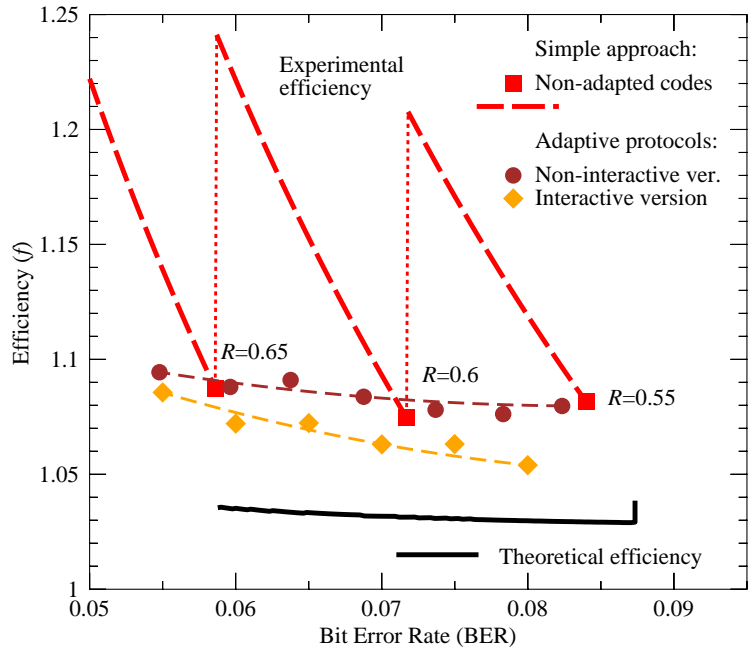}
\caption{Comparison between the efficiency obtained in the original rate adaptive protocol (non-interactive) using LDPC codes~\cite{Elkouss10} and the interactive version proposed here. It is also included the theoretical efficiency computed for the same code, and the efficiency calculated for LDPC codes without rate modulation~\cite{Elkouss09}.}
\label{fig:results}
\end{figure}

Finally, Fig.~\ref{fig:results} shows the reconciliation efficiency curves, calculated according to Eq.~\ref{eq:efficiency}, but obtained with three different approaches: i) the protocol proposed here where the maximum number of allowed rounds has been increased to 20, ii) the original non-interactive protocol~\cite{Elkouss10}, and iii) an ensemble of LDPC codes for different rates.

\section{Conclusions}
\label{sec:conclusions}

In this paper we have studied an interactive information reconciliation protocol. The protocol has been analysed empirically and the different trade-offs in terms of decoding complexity, interactivity and efficiency have been described.

The protocol has several advantages. The interactive nature of the reconciliation improves the decoding process, whenever the decoding fails, a fraction of the punctured symbols is revealed thus allowing for virtually zero error rate after decoding. The adaptive characteristic of the protocol allows to skip measuring the error rate on the channel. Several applications can boost their performance if this step is skipped: an important example is secret key distillation in QKD protocols. In this context the error rate is measured by publicly showing a subset of the sequences and discarding the shown symbols, the elimination of this step allows the parties to distill a significantly higher secret key rate.

The protocol presented on this paper can find a broad range of applications as it allows the parties to achieve reconciliation efficiencies as low as desired and, in several scenarios, to avoid the waste of a relevant part of the sequence for sampling purposes.

\section*{Acknowledgment}

This work has been partially supported by the project Quantum Information Technologies in Madrid\footnote{http://www.quitemad.org} (QUITEMAD), Project P2009/ESP-1594, \textit{Comunidad Aut\'onoma de Madrid}.

The authors would like to thank the assistance and computer resources provided by \textit{Centro de Supercomputaci\'on y Visualizaci\'on de Madrid}\footnote{http://www.cesvima.upm.es} (CeSViMa).

\bibliographystyle{IEEEtran}
\bibliography{istc2010}

\vfill

\end{document}